\documentclass{aa} 
\bibpunct{(}{)}{;}{a}{}{,} 

\usepackage{natbib}
\usepackage{caption}
\usepackage{float} 
\usepackage{subcaption}
\usepackage{graphicx}
\usepackage{txfonts}
\usepackage{verbatim}
\usepackage{booktabs}

\begin{document} 

   \title{Concurrent infall of satellites} 

   \subtitle{Collective effects changing the overall picture}

   \author{A. Trelles\inst{1}, O. Valenzuela\inst{1}, S. Roca-F\'abrega\inst{2} \and H. Vel\'azquez\inst{1} 
          }

   \institute{Universidad Nacional Autónoma de México, Instituto de Astronomía, AP 70-264, CDMX  04510, México\\
              \email{jtrelles@astro.unam.mx}
         \and
         Departamento de F\'isica de la Tierra y Astrof\'isica and IPARCOS, Facultad de Ciencias F\'isicas, Plaza Ciencias, 1, Madrid, E-28040, Spain\\
             }

  \date{Received March 3, 2022; accepted in Astronomy $\&$ Astrophysics in September 26, 2022}

 
  \abstract
{In recent years, high-quality observational data have allowed researchers to undertake an extensive analysis of the orbit of several Milky Way satellite galaxies, with the aim to constrain its accretion history. Although various research groups have studied the orbital decay of a satellite galaxy embedded inside a dark matter halo, a large variety of new physical processes have been proven to play an important role in this process, but its full scope  not yet understood.}
{Our goal is to assess whether the orbital history of a satellite galaxy remains unchanged during a concurrent sinking. For this purpose, we analyzed the impact that the internal structure of the satellites and their spatial distribution inside the host halo may have on the concurrent sinking process due to both mass loss and the combined effect of self-friction -- as processes that have not been studied before for the concurrent sinking of satellites.}
{We set up a set of N-body simulations that includes multiple satellites that are sinking simultaneously into a host halo and we compared them with models that include a single satellite.}
{The main result of our work is that the satellite's accretion history differs from the classical isolated view when we consider the collective effects. Furthermore, the accretion history of each satellite strongly depends on the initial configuration, the number of satellites present in the halo at the time of infall, and the internal properties of each satellite. We observe that compact satellites in a flat configuration fall slower than extended satellites that have lost mass, showing a non-reported behavior of dynamical self-friction; the latter is reinforced by analytical expressions that describe the orbital decay through different approaches for the dynamical friction, including (or not) the mass loss and radial dependence of the satellite. In particular, we find that such effects are maximized when satellites are located in a flat  configuration. Here, we show that in a flat configuration similar to the observed vast polar structure, deviations in the apocenters can be of about 30$\%$ with respect to the isolated case, and up to 50$\%$ on the eccentricities.} 
{Overall, we conclude that ignoring the collective effects produced by the concurrent sinking of satellite galaxies may lead to large errors in the determination of the merger progenitor properties, making it considerably more challenging to trace back the accretion event. Timing constraints on host density profile may be modified by the effects discussed in this paper.}

               
   \maketitle

\section{Introduction} 
\label{sec:intro}
The streamers of the Milky Way are true fossil records providing information about their formation and evolution history \citep{2018ApJ...863...89S,2020ApJ...893...48N,2013ApJ...773L...4V,2010ApJ...714..229L}.  Dynamical friction (DF, hereafter) is arguably the most critical physical process driving the sinking of galactic systems in a gravitational potential well. It was first studied by \citet{1943ApJ....97..255C} (CH, hereafter), based on a simple model that  assumes infinite and homogeneous media, while simultaneously capturing the dominant physical effects. 

Other authors have proposed new strategies to take into account other factors involved on the DF problem, beyond the Chandrasekhar approximation. Among them, the most relevant are those considering a more realistic host system, that is, with a finite size and a radially dependent density profile \citep{1977MNRAS.181..735B,2018arXiv180609591B}, as well as those considering that realistic host systems such as those in N-body models can be well described by a simple Coulombian logarithm with spatial or mass dependences  \citep{2003ApJ...582..196H,2005A&A...431..861J,1999ApJ...525..720C,2008MNRAS.383...93B}. All the new models have allowed researchers to recover and study the slowdown of the DF efficiency, while extending the realism of the system. Finally, internal changes in the perturber may also affect its sinking process and this should be taken into account when studying the accretion of satellite systems into central galaxy halos \citep{1998MNRAS.294..465D}. 

All the aforementioned studies on simplified models of satellites or globular clusters sinking into central galaxies can now be used to better understand the formation, evolution, and properties of local galaxies including the Milky Way. Far from the idealized models presented above where a single object sinks into a central halo, observations and simulations show that central halos suffer a continuous and concurrent infall of satellites \citep[e.g.,][]{2012MNRAS.425..231R}. The collective effects of the concurrent infall of satellites is a process that is pending further study. In this work we address it, for the first time, in a systematic way, showing that the sinking orbit of each individual satellite is sensitive to the relative configuration of all satellites inside the main halo and also to their internal structure. In particular, we have given some attention to an specific configuration similar to the recent plane such as the configuration that has been found for several galactic systems, known as the vast polar structure (\citet{1976MNRAS.174..695L,1976RGOB..182..241K}, VPOS, hereafter).

\section{Simulations and methodology}
\label{sec:method}
\subsection{Methodology}
\label{sec:methodology}
To carry out the research presented here, we generated a series of N-body experiments that include a spherical distribution of particles initially in dynamical equilibrium (central system) and one or various perturbers. We compare the evolution of these N-body systems (NB, herafter) with the results from semianalytical models (SA, hereafter) as described below.

\subsection{Semianalytical models} 
\label{sec:semiamodel}
Our semianalytical fiducial models assume a cuspy dark matter halo profile \citep{1996ApJ...462..563N} for the host (NFW, hereafter), the satellites are represented by a rigid Plummer model and we include the CH dynamical friction. Additionally, we incorporated the mass loss of the sinking satellite following \citet{1999ApJ...516..530K}. We also took into account a non-homogeneous density profile for the host by incorporating a radially dependent Coulombian logarithm. For simplicity, we chose the strategy proposed by \citet{2003ApJ...582..196H} (see their equation 2). We used these semianalytical experiments as control models to guarantee accuracy in the simulations in the regime in which SA and NB simulations are comparable and to identify the relevant physical processes from the ones included in the SA models.

\begin{table}
\caption{Initial parameters of the host and the satellite models used in the N-body simulations.}
\centering
\begin{tabular}{|c|c|c|c|c|}
\hline
&Host&Sat. comp.&Sat. Intermed.&Sat. ext.\\
\hline
M$_{200}$/M$_{\odot}$&1.2$\times$10$^{12}$& 10$^{10}$&10$^{9}-$10$^{10}$&10$^{10}$\\
N$_p$&2$\times$$10^{6}$&1.3$\times$$10^{4}$&1.3$- $13$\times$$10^{3}$&1.3$\times$$10^{3}$\\
r$_{s}$[kpc]&15&0.5&3&10\\
r$_{1/2}$[kpc]&20&1.5&5.5&18\\
$M_{1/2}$/M$_{\odot}$&6$\times$10$^{12}$&5$\times$10$^{11}$ &0.5-5$\times$10$^{11}$&5$\times$10$^{11}$\\
R$_{tr}$ [kpc]&200&1&5&25\\
$\epsilon$ [kpc]&0.3&0.3&0.2$-$0.4&0.6\\
\hline
\end{tabular}
\label{table:1}
\tablefoot{The satellite models include rigid, compact live (globular cluster type), intermediate live (dwarf galaxy type), and extended live (early type). From top to the bottom: Total mass (M$_{200}$), the number of particles (N$_p$), the scale radius (r$_{s}$) of the NFW density profile, the half mass radius and half mass (r$_{1/2}$,M$_{1/2}$), the truncation radius (R$_{tr}$), and the smoothing length ($\epsilon$). For rigid satellite M$_{200}$/M$_{\odot}$ = 1$\times$10$^{10}$, $\epsilon$ = 0.6 kpc
}
\end{table}

\subsection{N-body simulations}
\label{sec:nbody}
The set of NB simulations we used here include a central system (the host, hereafter) that is modeled as a simple dark matter halo with a NFW density profile and composed of particles that interchange energy (hereafter, live systems), plus one or several perturbers (satellites, hereafter). Our suite of models includes systems with rigid (softened Plummer), and either compact live, intermediate live, or extended live (see table~\ref{table:1}). All N-body models are labeled as NB$_{x}$ where \textit{x=r,c,e} and corresponds to a Plummer rigid satellite, a self-consistent compact satellite and an extended NFW satellite, respectively. In all models, the host has a mass of M$_{200}$=1.2$\times$10$^{12}$M$_{\odot}$ and a total of 2$\times$10$^{6}$ particles. The number of particles in the satellites varies to keep the mass ratio of the host  same to the satellite particles. The softening of the host and satellite particles is taken as the average particle separation within the scale radius (see Table~\ref{table:1}). The initial conditions were obtained using a NFW model, which includes a cutoff in the form of an hyperbolic secant weight function, and depends on the truncation radius $R_{tr}$ parameter. To create the initial conditions we used the subroutine named "mkhalo" from \citet{2005MNRAS.363.1057D} that can be found in the NEMO toolkit package \citep{2010ascl.soft10051B}. We used the public N-body code "gyrfalcon" which is a tree code with complexity $\sim$O(N), combined with a fast multipole method, (FMM; \citet{2014ascl.soft02031D}) to evolve the aforementioned initial conditions for 10$-$20~Gyr. To track the path of live satellites in the NB simulations, we used the Rockstar halo finder \citep{2013ApJ...762..109B}.

\begin{figure}
     \centering
     \begin{tabular}{cc}
        \includegraphics[width=0.4\textwidth,angle=0]{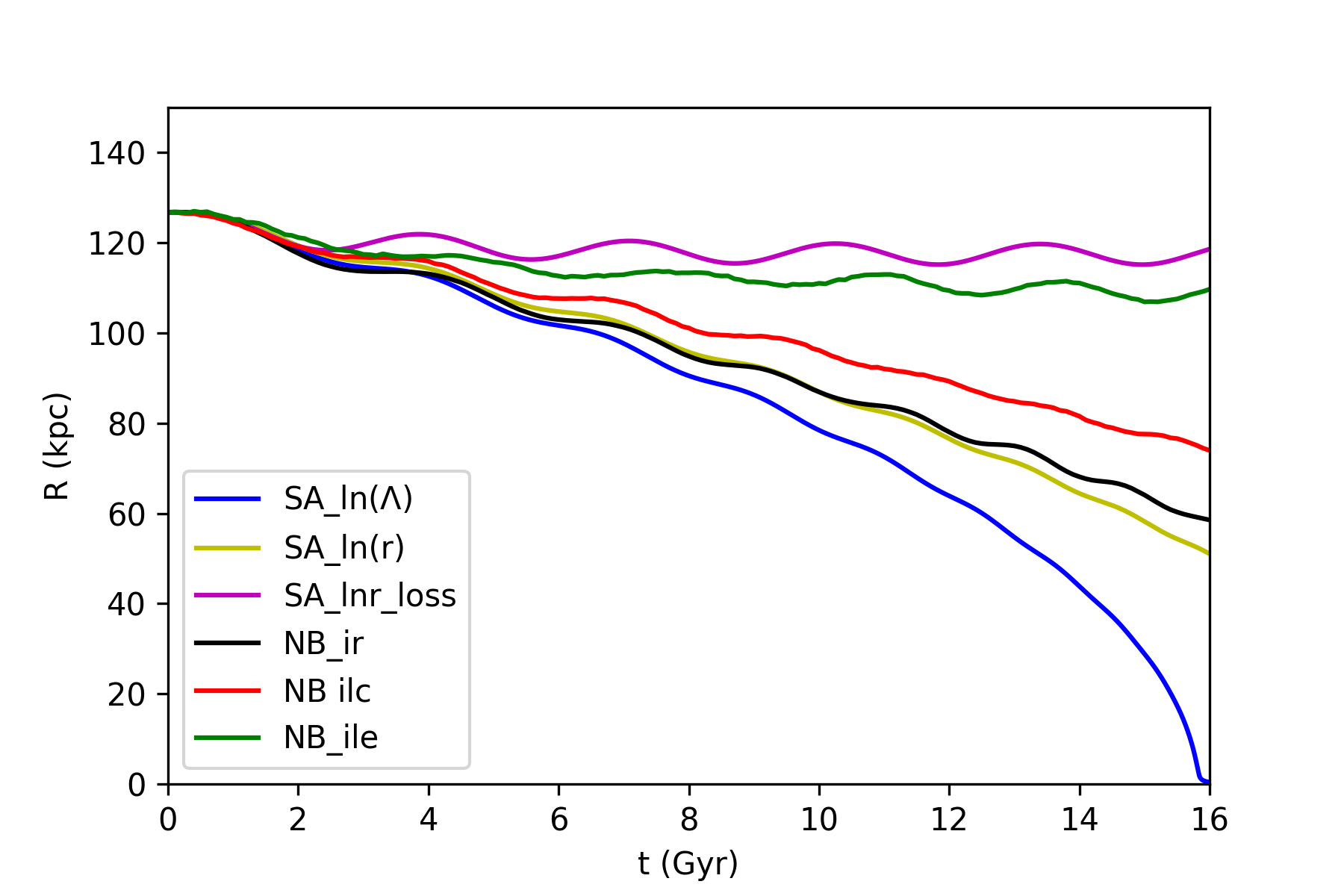}
     \end{tabular}
\caption{Sinking of single satellites in a MW-mass halo.  Blue line: SA fiducial model as a Coulombian logarithm with no spatial dependence;  Yellow: SA using a  Coulombian logarithm with radial dependence;  Magenta: SA using a Coulombian logarithm with radial dependence and a mass loss; Black: NB model with a satellite represented with a rigid softened Plummer density profile (rigid system, hereafter), $\Lambda = \frac{b_{max}}{b_{90^{o}}}\approx5 $, following \citet{1987gady.book.....B} eq. 8.1b ; Red: NB model including a satellite with a compact NFW density profile; Green: NB model including a satellite with an extended NFW density profile (for more details on the models see Sect.~\ref{sec:results1} and table\ref{table:1}). The initial condition for the satellite in the models is r = 126 kpc, v$_{c}$ = 183 cm/s.} 
\label{fig:fit} 
\end{figure}

\begin{figure}
     \centering
        \includegraphics[width=.28\textwidth,height=.7\textwidth]{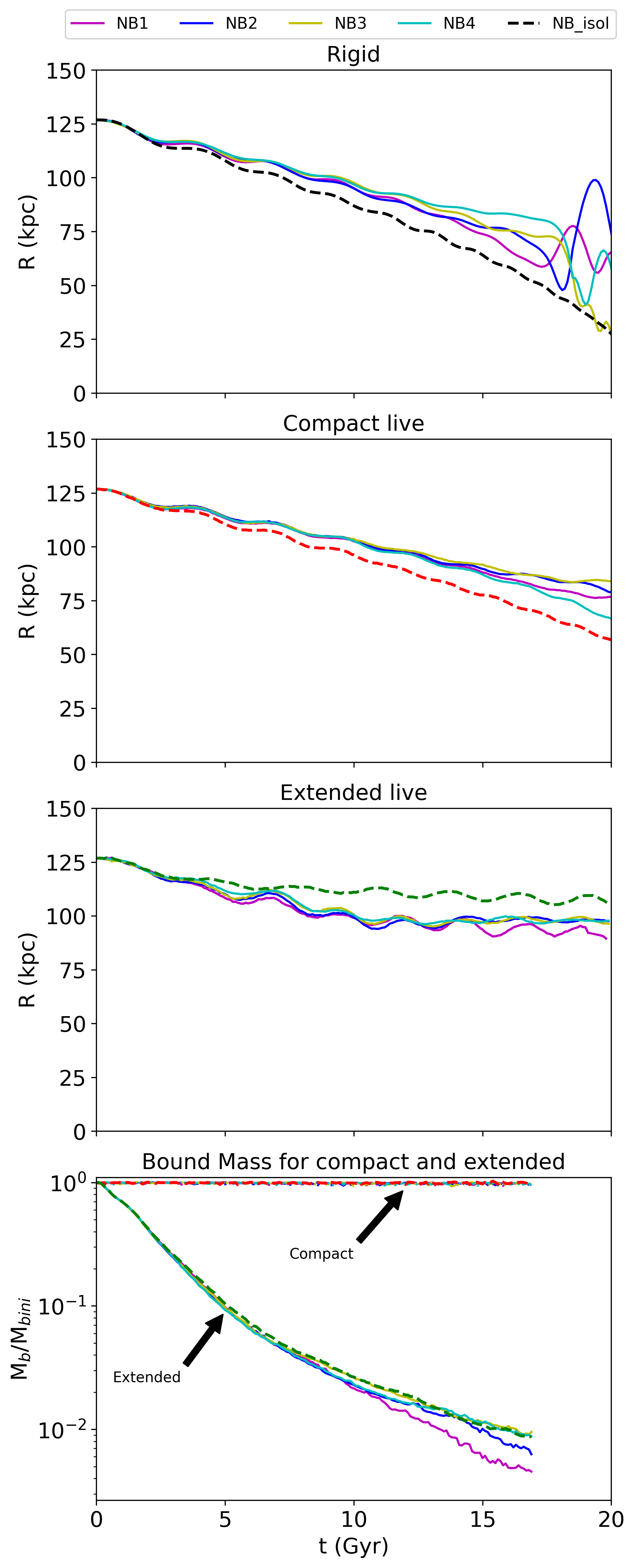}
\caption{NB models of a MW-mass halo host with four sinking satellites in the same orbital plane and in a fully symmetric configuration (see Figure~\ref{fig:yt}), and with initial speeds corresponding to circular orbits (upper panels). From top to the bottom, satellites simulated as a rigid smoothed Plummer density distribution, as a compact NFW, and as an extended NFW. For comparison purposes, we include the sinking process of a single satellite in all panels (dashed lines). Mass evolution of the four compact and extended satellites in their collective sinking process (bottom panel). The dashed lines show the same process when only one satellite sinks. Satellites start at r = 126 kpc and with $v_{c}$ = 183 km/s} 
\label{fig:foursate}  
\end{figure}

\begin{figure}
     \centering
        \includegraphics[width=0.52\textwidth,height=0.42\textwidth,angle=0]{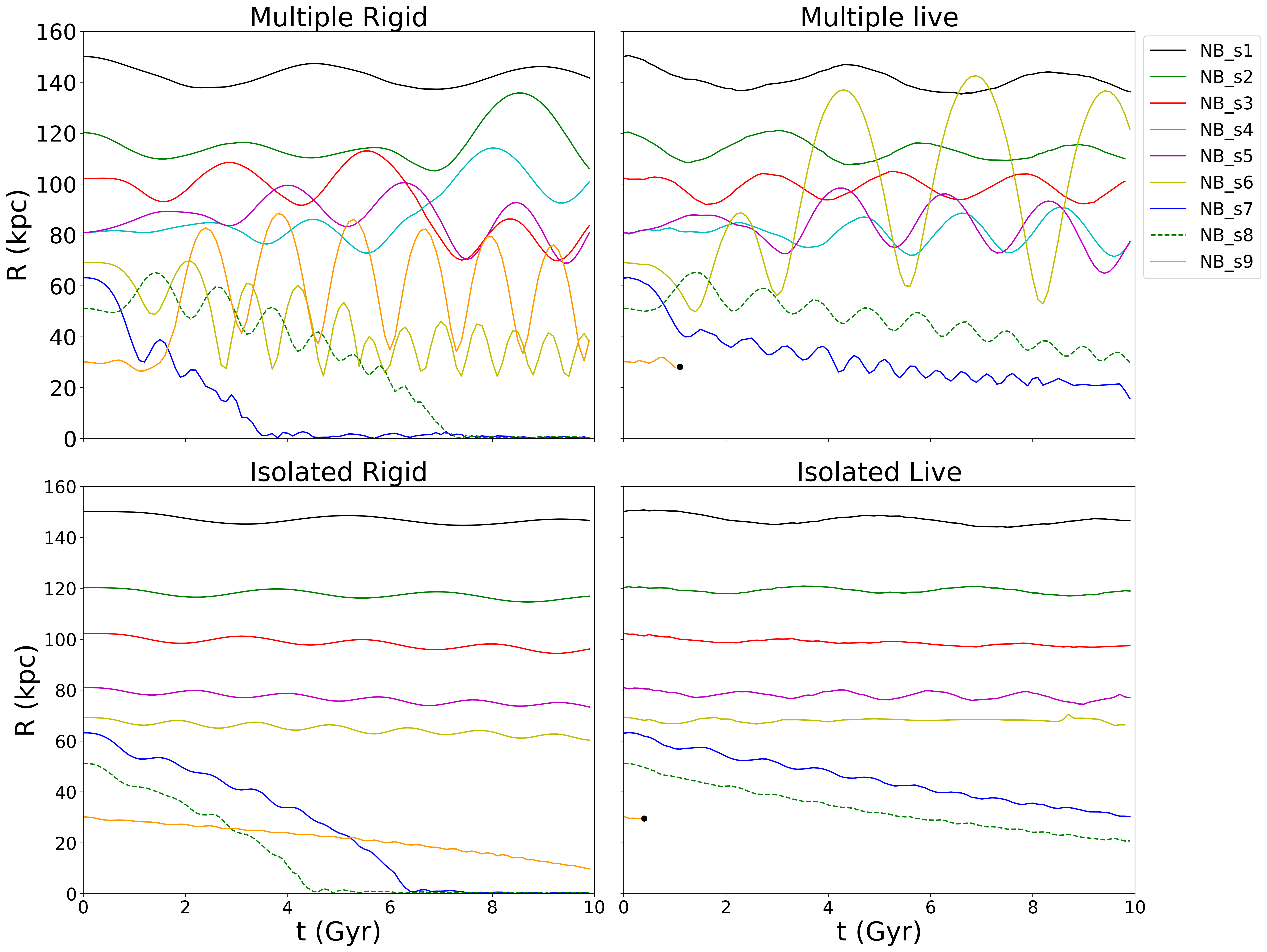}
\caption{NB models of a MW-mass halo host with nine sinking satellites located in the same orbital plane but falling from different angular and radial positions (see Figure~\ref{fig:yt}, bottom panels). All satellites have an initial speed corresponding to the one of a circular orbit. {\it Top-left panel:} Simultaneous infall of the nine satellites simulated as a rigid smoothed Plummer density distribution. {\it Top-right panel:} Similar to the top-left panel, but satellites are now described by an "intermediate" compact NFW density distribution (see column 4 in table~\ref{table:1}). {\it Bottom-left panel:} Sinking of each one of the rigid satellites in the top-left panel, individually, meaning that no other perturber is present in the halo. {\it Bottom-right panel:} Similar to the bottom-left panel but for the "intermediate" compact satellites. In each of the bottom panels we show eight independent experiments, where the initial mass profile of the host halo has been modified to include the mass of the remaining non-simulated satellites at its initial radial location. We excluded the cyan line case because it is the same as the magenta case. Black points indicate where the Rockstar halo finder cannot find the satellite because it lies in the central part of the host, See Table~\ref{table:2} for the initial conditions.}
\label{fig:copla_case}
\end{figure}

\begin{table}
\caption{Initial parameters for the satellites in the coplanar experiments.}
\centering
\begin{tabular}{|c|c|c|c|}
\hline
NB & r (kpc) & v$_{t}$ (10 km/s) & M$_{\odot}$\\
\hline
S1 & 30.16 &  21.68 & 10$^{9}$\\
\hline
S2 & 51.07 & 21.26 & 10$^{10}$\\
\hline
S3 & 63.18 & 20.95 &  10$^{10}$\\
\hline
S4 &  69.21 & 20.84 &  10$^{9}$\\
\hline
S5 & 81.07  & 20.31 &  10$^{9}$\\
\hline
S6 &  81.07 & 20.31 &  10$^{9}$\\
\hline
S7 & 102.06 & 19.44 &  10$^{9}$\\
\hline
S8 & 120.06 & 18.75 & 10$^{9}$ \\
\hline
S9 & 150.05 & 17.71 &  10$^{9}$\\
\hline
\end{tabular}
\tablefoot{From left to right: the radial distance, the tangential velocity to that radio, and the initial mass for the satellites in experiments from Figure~\ref{fig:copla_case}. The corresponding live satellites in this experiments corresponds to the intermediate compact satellite that appears in  column 4 in Table~\ref{table:1} and these satellites lie on a plane.}
\label{table:2}
\end{table}

\section{Results}
\label{sec:results}
\subsection{Dynamical Friction: Single satellite sinking using N-body and semianalytical models}\label{sec:results1}
In Figure \ref{fig:fit}, we present the comparison between our SA and NB simulations. In this set of six models, the complexity of the simulations is gradually increased. In the first three models, we used a SA approach where DF is first modeled the classical Chandrasekhar description using a constant Coulomb logarithm and no mass loss (blue); then we used a spatially dependent Coulomb logarithm and no mass loss (yellow); and in the last run (magenta), we used a spatially dependent Coulomb logarithm and including a mass loss equal to the one corresponding rigid satellite NB experiment (solid-black line). The last three models are NB models consisting of a live host galaxy and a single satellite that is modeled as rigid (black), compact and live (red), and extended and live (green). This first set of six models is relevant in our work as it illustrates agreement between the SA and NB models when the former includes the physical processes that play a dominant role in the NB simulations (e.g., yellow vs. black and pink vs. green lines in Figure~\ref{fig:fit}). They serve as our reference models to compare with the multiple satellite configurations with the aim to identify (or search) for dynamical processes that are not taken into account.

\subsection{Concurrent satellite sinking and internal structure}\label{sec:concurrent}
The multiple satellite case is complicated because, in addition to the standard processes discussed in the previous section, satellites interact among them as well as with the corresponding halo response and the mass lost by the satellites. These effects affect each other and their relative role in the sinking rate can not be trivially separated. 
As our main objective is to study differences between isolated models and models with many perturbers, we set a spatial configuration that, after performing many experiments with random positions for the satellites, we found it maximizes the collective effects. This configuration is a fully symmetric and flat (coplanar) distribution (see Figure~\ref{fig:yt}, upper panels). \\
In Figure~\ref{fig:foursate}, we show the results obtained after analyzing three NB models with four rigid, compact live, and extended live satellites, respectively, sinking simultaneously through a live MW-mass halo (pink, blue, cyan, and yellow solid lines), and we compare them with the corresponding experiment of a single live satellite (black, red, and green dashed lines). In the top panel, we show that rigid Plummer satellites sink at a slower rate than the single satellite case (black-dashed). In the second panel, we show the sinking of four compact NFW live satellites and we obtain a similar result as before, satellites sink at a slower rate in the collective scenario than in the single satellite case (red-dashed). However, in agreement with the results presented in Section~\ref{sec:results1}, we observe differences between the rigid and the compact live scenarios; as the mass loss is almost negligible in the compact live scenario (as detailed in the discussion below), the only source of these differences can be the internal structure of the live satellite that can work as a sink or source of energy and angular momentum \citep{1998MNRAS.294..465D}. In the third panel, we show the trajectories of satellites when these are modeled as live extended NFW; we see that extended satellites sink faster in the concurrent scenario than when they sink in isolation (green-dashed). Although it may seem unexpected the result can be explained by analyzing the mass loss suffered by satellites. In the bottom panel, we show the mass loss for the compact and extended satellites. As expected, this figure confirms that extended satellites in the collective scenario suffer higher mass loss in comparison with the compact ones. If we focus only on the extended satellites, we see that the ones sinking faster also suffer higher mass loss than the isolated ones, which is unexpected based on Chandrasekhar formula. We hypothesize that the interaction of the stripped material from these satellites among themselves and the others increases the self-friction and, thus, the sinking rate. \\
We assert that this contrasting behavior between the single satellite and the four extended ones shows an aspect of self-friction that has not been studied before. Therefore, the combined effect of stripped material of nearby satellites may increase their sinking speed. This may suggest that our assumption of a coplanar symmetric configuration precludes our conclusions from proving relevant in real galaxies. Although it is not infrequent to find satellite galaxies sinking in small groups, in pursuit of the answer to this question, we study a more realistic case, described in the next section.

\subsection{Connection with realistic systems: Vast polar structure}\label{sec:realistic}
\begin{figure}
     \centering
\includegraphics[width=0.49\textwidth,height=0.6\textwidth,angle=0]{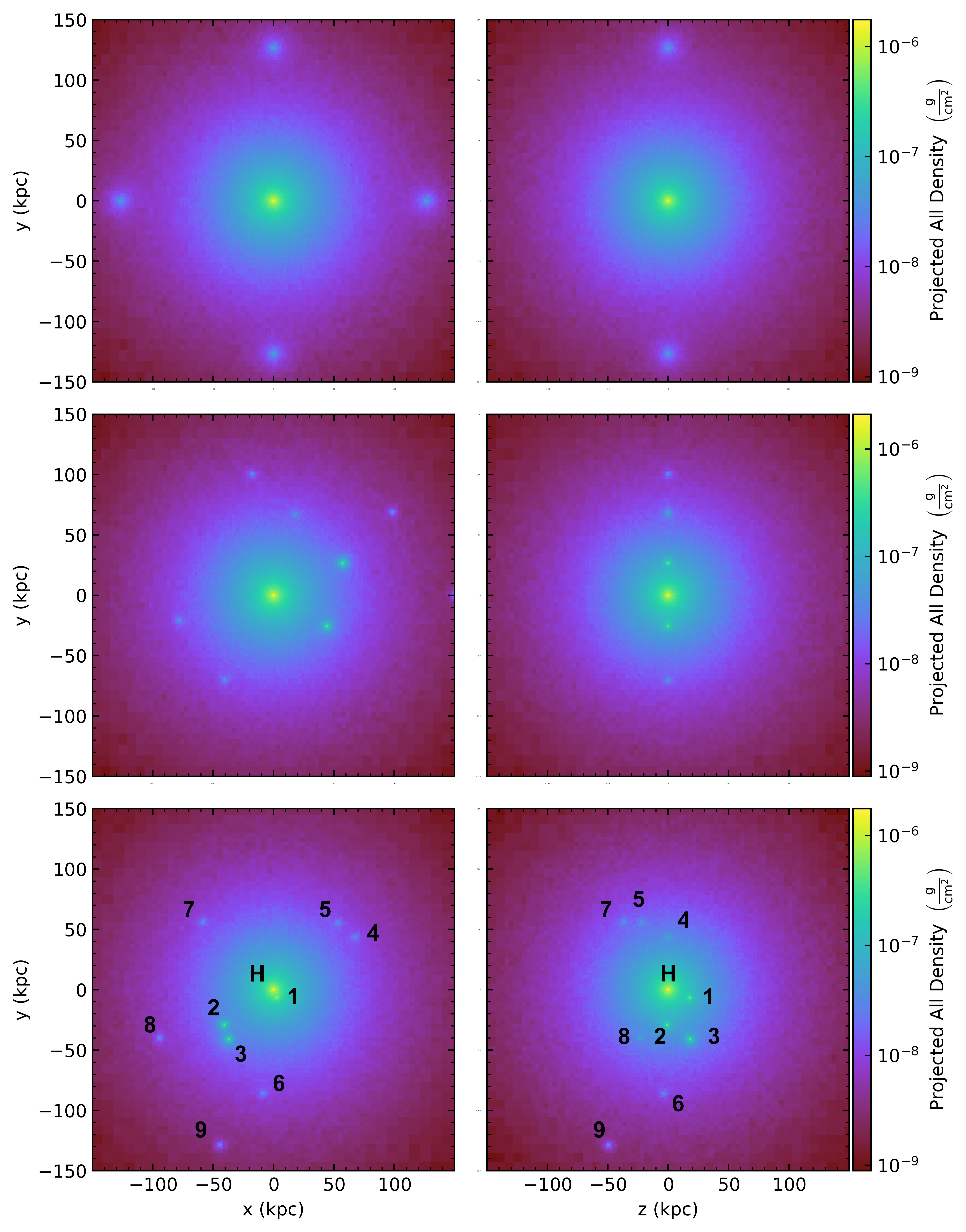}
\caption{Density plots showing the satellites' initial positions in the NB models. {\it Top panels}: Face-on (left) and edge-on (right) views of the fully symmetric model with four satellites of the same mass that follows an extended NFW density profile. {\it Central panels}: Face-on (left) and edge-on (right) views of a fiducial model, where satellites lie in a plane as an approximation model to the VPS plane.  {\it Bottom panels}: Face-on (left) and edge-on (right) views of one of the models in our N-body ensemble that follows the distribution of the satellites in the MW and the VPS plane structure; see \ref{section:robustness} for details. Labels are consistent with the results showed in figure \ref{fig:std_apo_peri}, where H stands for the host's central density.}
\label{fig:yt}
\end{figure}

In the previous section, we show that the sinking history of four satellites, initially located in a symmetric and coplanar configuration, dramatically differs from one that involves solely an individual satellite. While it may appear idealistic, such a coplanar symmetric configuration somehow is in agreement with recent studies reporting the so-called vast polar plane of satellite galaxies (VPOS) in the local group and other galaxies \citep{2005A&A...431..517K,2012MNRAS.423.1109P}. To better study the connection between our results and the VPOS problem, in this section we analyze experiments  considering nine satellites in a plane configuration resembling the VPOS, using the relative orientations presented in \citet{2020MNRAS.491.3042P} (see also Figure~\ref{fig:yt}, bottom panels). It is important to clarify that we do not pretend to develop a realistic MW+satellite model; rather, we built a test system with a similar flatness as the VPOS in order to test the relevance of our former results. Additionally, we ran some experiments where the satellites were located at different distances out of the plane and  we also used both the relative positions and velocities presented in \citet{2020MNRAS.491.3042P}. The results are consistent with our general conclusions, with their statistical properties are discussed in the next section. However, a detailed discussion will be presented in a forthcoming work. In the following, we analyze two experiments, one with nine rigid satellites and another with nine "intermediate"  ones (see column 4 in Table~\ref{table:1} and its description). In these experiments, the two satellites marked as blue and green dashed line,  have an initial mass of $10^{10} M_{\odot}$, while the rest of the others have a mass of $10^{9} M_{\odot}$.

The results are shown in Figure~\ref{fig:copla_case}. As a reference, we show the isolated sinking history of each satellite in the lower panels, both for the rigid (bottom-left) and the live (bottom-right) representations. It is important to notice that in the reference models (isolated satellites) that appear in lower panels of Figure~\ref{fig:copla_case},
we update the initial mass profile of the halo to include the mass due to the more internal satellites than the respective isolated case and that would be present in the case of multiple satellites.
 
Our analysis of both scenarios (rigid and live) shows that most of the trajectories of the satellites are affected when including the collective effects. We observe a wide range of variations, from subtle changes in the orbits of outermost satellites (s1, black curves) to more extreme ones such as what is seen in the s6 or s3, and also complex interactions between perturbers. However, the scope of this experiment is not to analyze the sinking of each individual satellite, but to demonstrate that collective effects change the orbits of the satellites in a way that would make it challenging to recover the infall history of satellite galaxies based on current observations. This is true for such compact systems as globular clusters (i.e., similar to rigid satellites) and for satellites such as Sagittarius (i.e., similar to the "intermediate" live satellites). Only systems in the outermost regions of the host halo suffer minor perturbations, and so their orbits can be easily tracked back in time.

\begin{figure}
     \centering
     \begin{tabular}{cc}
    \includegraphics[width=0.4\textwidth,height=0.55\textwidth,angle=0]{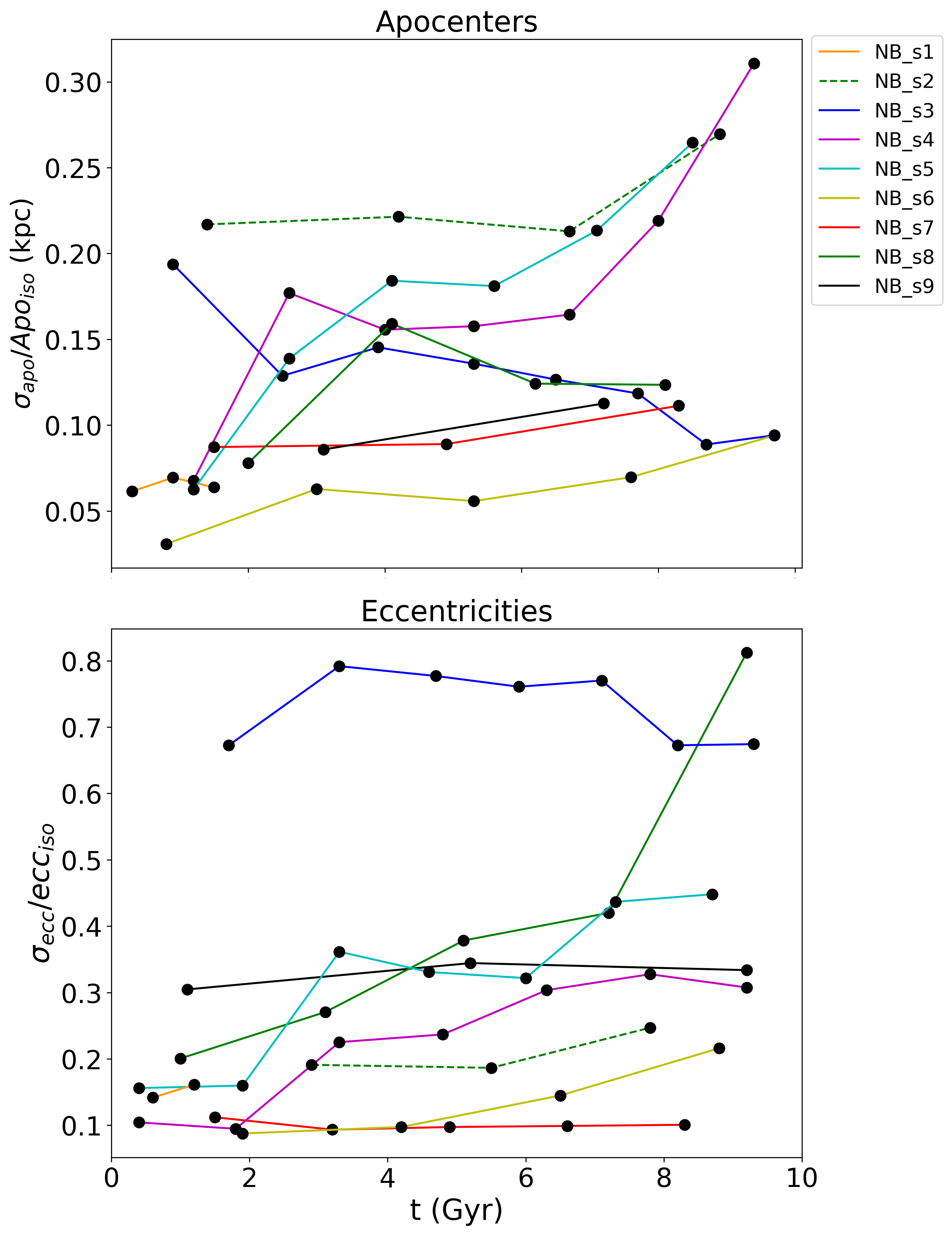}
     \end{tabular}
\caption{ Standard deviation in the apocenters (top) and eccentricities (bottom) when comparing results from our ensemble of N-body models and the isolated case (see section  \ref{section:robustness}).  Eccentricity is normalized with the initial eccentricity of the isolated case for every satellite.}
\label{fig:std_apo_peri}
\end{figure}

\subsection{Exploring the statistical robustness}
\label{section:robustness}
To quantify the statistical stability of our results, we generated an ensemble of 20 N-Body models of a host MW-size galactic halo and satellites, as shown in Table~\ref{table:2} from \citet{2020MNRAS.491.3042P}. In each model of the ensemble, we changed satellites' positions and velocities, randomly generated within the ranges defined by the observational uncertainties \citep{2020MNRAS.491.3042P,2019MNRAS.486.2679R,2020MNRAS.499..804G}; in addition, we followed a Gaussian distribution, so we could estimate the effect of changes in the initial orbital parameters. We note that our models do not include the Leo I and Leo II satellites due to their large uncertainties in GAIA DR2 proper motions.\\
We studied the orbital parameters of each satellite in each model of our ensemble. In particular, we analyzed the time and distance to the center of mass during apocenters and pericenters to get also the eccentricity of each orbit. For each satellite we tracked the "rms" changes of apocenters and we even computed the rms deviation of eccentricity with respect to the isolated case. In addition, we carried out an extensive study of the number and intensity of close encounters between satellites. 

\begin{figure}
\centering
\begin{tabular}{cc}
\includegraphics[width=0.44\textwidth,height=0.3\textwidth,angle=0]{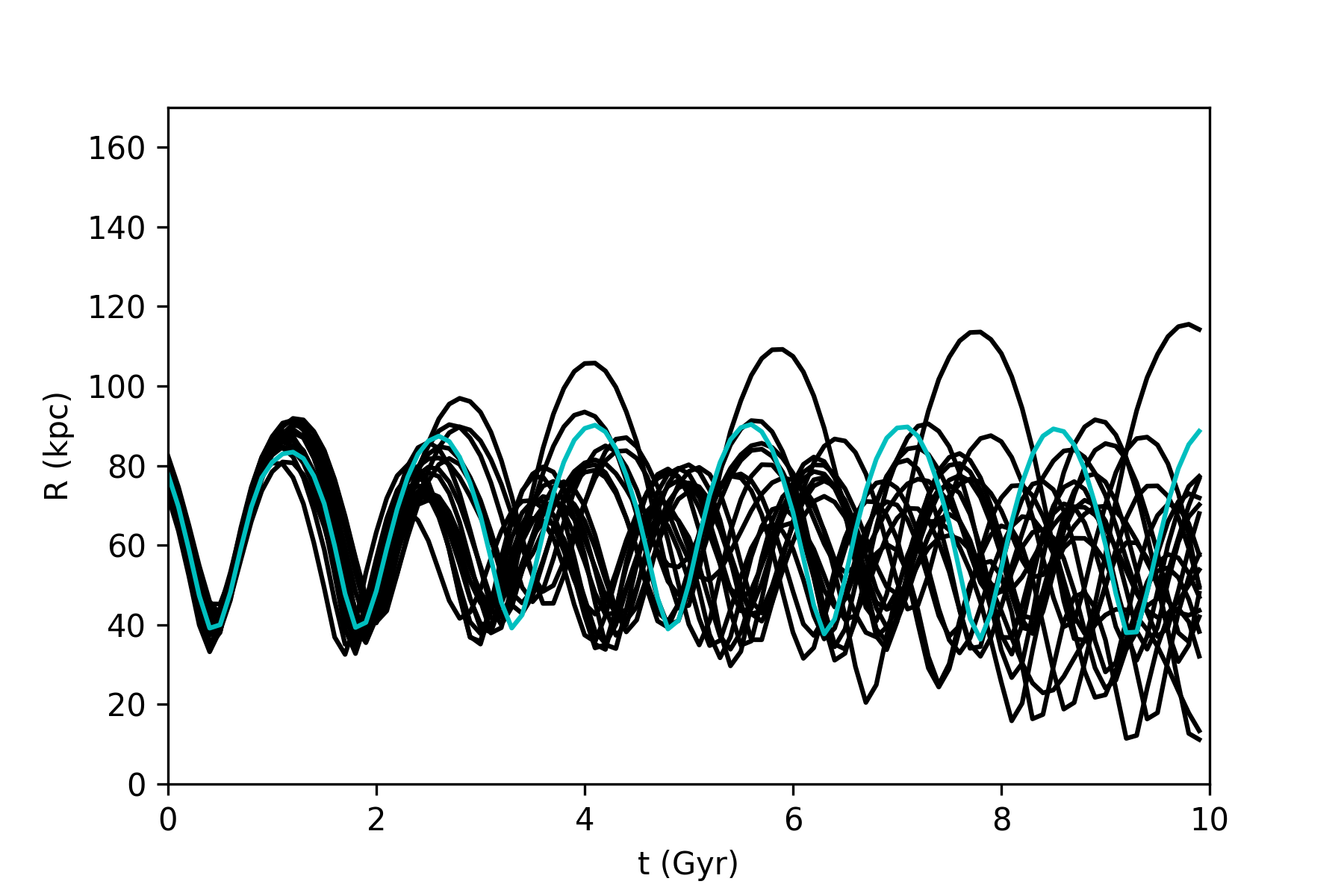}\\
        \includegraphics[width=0.42\textwidth,height=0.3\textwidth]{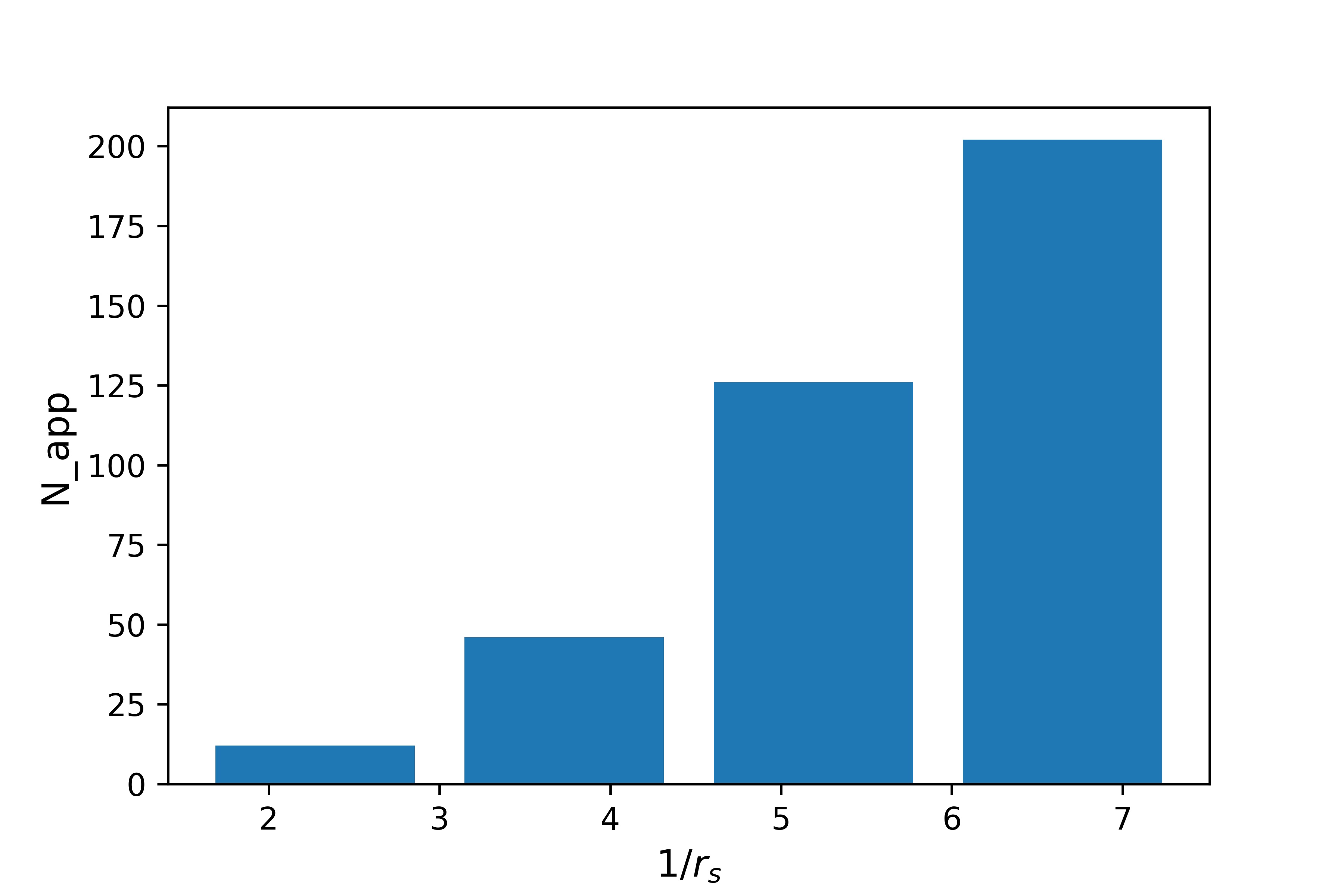}
        \end{tabular}
      \caption{Comparison of the orbit of a particular satellite (S5/6 according to table~\ref{table:2} nomenclature) when run in isolation (cyan line) or in each one of the 20 N-body models of our ensemble, shown in the top panel (black solid lines). This figure shows the scatter between realizations that we quantified in Figure \ref{fig:std_apo_peri}, for one of the satellites. Bottom panel shows the histogram of distances of all close encounters between two satellites in the full ensemble of N-body models.}
      \label{fig:examples}
\end{figure}
\section{Discussion and conclusions}\label{sec:conclusions}

In this paper, we carry out the first systematic study of the effects of concurrent multiple sinking perturbers. More specifically, we analyze how such effects produce changes on the orbits of satellite galaxies.  In conclusion, we find clear differences between the concurrent and single-satellite scenarios, which would be partially explained by the fact that the global (host + satellites) potential is time-dependent. Although close encounters between satellites may happen, we discard them as a dominant mechanism. When looking at the upper panel of Figure~\ref{fig:examples}, we see that none of the ensemble orbits agree with the isolated case regardless of the presence or absence of a close encounter and the orbital drift is increasing with time. However slingshot effects still may happen, combined with the satellites interaction with the local and global halo density response, as well as interactions with the combined stripped material or dynamical self-friction (Figure~\ref{fig:foursate}). We enumerate our more relevant results in the following. 

\begin{itemize}
\item The sinking process of satellities, when including collective effects of several other perturbers, is generally differerent from that of an isolated satellite. The difference depends on the internal properties of the sinking satellites, the relative spatial and velocity configuration between them, the orbital parameters, and the mass ratio between the host and satellites, as well as the possible scatter between the different satellite masses. 

\item  In particular, in the case with multiple compact or rigid satellites, the sinking process is slower than for the isolated ones. In this scenario, most of the effects of  these interactions  (satellites and host density responses) go into satellite orbital energy, delaying the sinking process (see upper panels in Figure \ref{fig:foursate}). The former situation may be comparable with the compact globular clusters in dwarf spheroidal galaxies such as Fornax or Reticulus 2. However, it is necessary to carry out a future assessment of such case to test whether the rigid and compact satellite assumptions are suitable for globular clusters and also the effect of the relative globular cluster orbit orientation \citep[see also][]{  2011MNRAS.416.1181I}. 

\item For extended satellites in concurrent accretion, the mass loss is considerable. The sinking rate is faster in some situations, in comparison with the isolated case due to the combined effect of stripped material, revealing a new aspect of self-friction. Such a collective effect
leads to the self-friction amplitude sometimes rising higher than the $10-15\%$ quoted by \citet{2020MNRAS.495.4496M} (see lowest panel in Figure \ref{fig:foursate}).

\item  We found that the internal density profile of the satellites makes them more or less susceptible to stripping, affecting the satellite sinking rate in combination with the already mentioned combined effect of the stripped mass from multiple satellites enhancing the DF.
    
\item  We found that collective effects are more dramatic in such coplanar systems as the so-called Milky Way VPOS (see Figure~\ref{fig:copla_case}), this is because the mutual perturbations are more frequent, nearby, and stronger, making the sinking satellite history quite different to the isolated case. Although we did not pretend to simulate the VPOS formation and evolution, we can point out that the sinking rate can diverge from the results of single satellite infall, it may temporally reduce or reverse the sinking process. 

\item We built an ensemble of twenty models for the satellite system similar to the VPOS in order to quantify the rms change regarding the isolated case for satellite orbital parameters (apocenter and eccentricity) during concurrent evolutionary trajectories (see Figures \ref{fig:std_apo_peri} and \ref{fig:examples}). In general, the orbital evolution does not correspond to the isolated case, however, in many situations the difference is mostly quantitative, although the difference is systematically growing and in some cases the orbit is completely different. We also estimate that the fraction of satellite binary encounters closer than the characteristic halo radius (3 kpc) is 5$\%$ (as we see in the lower panel of Figure \ref{fig:examples}), indicating that the observed orbital drift is triggered not only for close encounters, but also for the satellites' interaction with the local and global halo density response, as well as the tides of the combined stripped material or dynamical self-friction.  In the future, we plan to explore all of these effects with a code that efficiently captures such physical processes \citep{2022A&A...663A..93A}.  

\item   As a general lesson, it is not always trivial to disentangle the history of a sinking satellite accretion by only taking the instantaneous position and velocity and a simple global potential, without consider the whole galaxy system evolution or at least the most relevant perturbers. Not considering the correct complexity may result in misleading conclusions about the mass or accretion time of the perturber, or about the overall shape and evolution of the host gravitational potential. 

\item  Thus, models that include the infall of a single satellite (e.g., Sagittarius or the LMC or SMC systems) have a limited predictive power. In this paper, we emphasize that collective effects need to be taken into account when studying the sinking of satellites -- not only in the Milky Way, but in other systems as well. Recently, other works such as \citep{2022MNRAS.512..739D}, reinforced the idea that satellite orbital history is affected by large uncertainties related with the host or most massive subhalos, the collective effects that we discuss here strengthen this picture.
\end{itemize}  

We emphasize again that not considering the collective effects in the orbital reconstruction may lead to misleading conclusions about the mass or accretion time of the perturber and the overall shape and evolution of the host gravitational potential. 

\begin{acknowledgements}

AT and OV acknowledges support from a DGAPA-UNAM grant IN112518, IG101222 and AG101620. HV acknowledges support from PAPIIT-UNAM under grant IN101918. The authors thank for the facilities of cluster computers: Atocatl and Miztli used for this projects. SFR acknowledges support from a Comunidad de Madrid postdoctoral fellowship under grant number 2017-T2/TIC-5592. His work has been supported by the Madrid Government under the Multiannual Agreement with UCM in the line Program to Stimulate Research for Young Doctors in the context of the VPRICIT under grant number PR65/19-22462. SRF also acknowledge financial support from the MINECO under grant number AYA2017-90589-REDT, RTI2018-096188-B-I00, and S2018/NMT-429. This research was partially supported through computational and human resources provided by the LAMOD UNAM project through the clusters Atocatl and Tochtli. LAMOD is a collaborative effort between the IA, ICN, and IQ institutes at UNAM. The authors thank the anonymous referee for the comments and support.
\end{acknowledgements}

%
%

\bibliographystyle{aa} 
\bibliography{bibfile}

\end{document}